\documentclass[pra,twocolumn,superscriptaddress,showpacs,preprintnumbers,amsmath,amssymb]{revtex4}

\usepackage{graphicx}
\usepackage{dcolumn}
\usepackage{bm}
\usepackage{ifpdf}
\usepackage{multirow}

\begin{document}

\bibliographystyle{apsrev} 


\title{Quantum Zeno and anti-Zeno effects induced by either frequent measurements, modulations, or a mix of them}

\author{Wenxian Zhang}
\affiliation{Key Laboratory of Micro and Nano Photonic Structures, Department of Optical Science and Engineering, Fudan University, Shanghai 200433, China}
\affiliation{Advanced Science Institute, RIKEN, Wako-shi, Saitama 351-0198, Japan}
\affiliation{Kavli Institute for Theoretical Physics China, CAS, Beijing 100190, China}

\author{A. G. Kofman}
\affiliation{Advanced Science Institute, RIKEN, Wako-shi, Saitama 351-0198, Japan}
\affiliation{Department of Physics, The University of Michigan, Ann Arbor, Michigan 48109-1040, USA}

\author{Jun Zhuang}
\affiliation{Key Laboratory of Micro and Nano Photonic Structures, Department of Optical Science and Engineering, Fudan University, Shanghai 200433, China}

\author{J. Q. You}
\affiliation{Department of Physics, Fudan University, Shanghai 200433, China}
\affiliation{Advanced Science Institute, RIKEN, Wako-shi, Saitama 351-0198, Japan}

\author{Franco Nori}
\affiliation{Advanced Science Institute, RIKEN, Wako-shi, Saitama 351-0198, Japan}
\affiliation{Department of Physics, The University of Michigan, Ann Arbor, Michigan 48109-1040, USA}

\date{\today}

\begin{abstract}
Using numerical calculations, we compare the collective transition probabilities of many spins in random magnetic fields, subject to either frequent projective measurements, frequent phase modulations, or a mix of modulations and measurements. For three different distribution functions (Gaussian, Lorentzian, and exponential) we consider here, the transition probability under frequent modulations is suppressed most if the pulse delay is short and the evolution time is larger than a critical value. Furthermore, decoherence freezing (with a transition rate equals to zero) occurs when there are frequent phase modulations, while the transition rate only decreases when there are frequent measurements and a mix of them, as the pulse delay approaches zero. In the large pulse-delay region, however, the transition probabilities under frequent modulations are enhanced more than those under either frequent measurements or a mix of modulations and measurements.
\end{abstract}


\maketitle

\section{Introduction}

Quantum coherence is of key importance in studying microscopic and mesoscopic quantum systems and in developing quantum devices, including quantum registers in quantum computing~\cite{Nielsen00, Ladd10, Stolze08, Schleich08, Buluta11} and spintronic devices~\cite{Zutic04}. Many methods have been developed to extend the coherence time of a quantum system, particularly those methods using the quantum Zeno effect via either frequent measurements or frequent modulations~\cite{Misra77, Cook88, Kofman00}.

Utilizing the quantum Zeno effect, the coherence time of a quantum system can be extended significantly in nuclear and electron spin systems~\cite{Haeberlen76, Zhang07b, Lee08, Yang08, Biercuk09, Du09}, trapped ions~\cite{Itano90, Itano91}, ultracold atomic Bose-Einstein condensates~\cite{Streed06, Ning11}, and other physical systems~\cite{Search00, Frishman01, Evers02, Frishman03, Fischer01, Kofman01, Khodjasteh05, Uhrig07, Wang08, Zhou09, Cao10, Cao11}. Two frequent (periodic) control methods are often employed: either projective measurements or strong modulations. A systematic comparison of these two methods was made by Facchi et al.~\cite{Facchi05}: By calculating the transition rates, they compared the quantum Zeno/anti-Zeno effect in a two-level system via three methods: either (i) frequent measurements, (ii) frequent modulations, or (iii) a strong coupling to an auxiliary state. They assumed that the transition probabilities have always an exponential form and thus a well-defined decay rate for short times.

\begin{figure}
\includegraphics[width=3in]{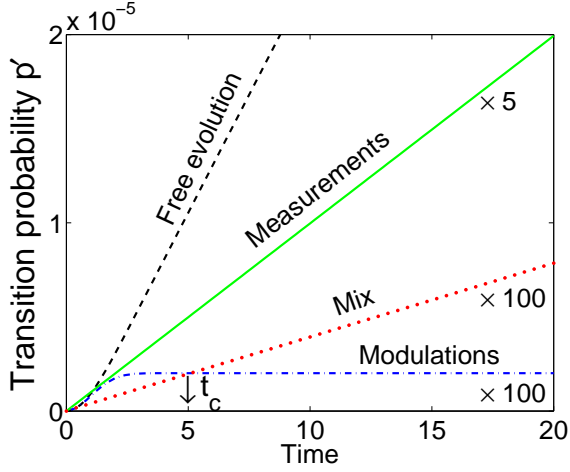}
\caption{(Color online) Comparison of transition probabilities $p'$ for free evolution (dashed line), under frequent measurements (solid line, multiplied by $5$), under frequent modulations (dash-dotted line, multiplied by $100$. Presented data are for even-number modulation pulses), and under a mix of measurements and modulations (dotted line, multiplied by $100$). Here the distribution has a Gaussian form [see Eq.~(\ref{eq:gaussian})] with $\omega_m = 0$ and $\Gamma=1$. The pulse delay is $\tau = 0.2$. The tiny arrow marks the critical time $t_c$ where the transition probabilities are the same under the modulations and under the mix of modulations and measurements. Frequent measurements and the mix method suppress the transition rate while frequent modulations freeze the transition probability.}
\label{fig:gtaus}   
\end{figure}

Zhang et al.~\cite{Zhang07a, Zhang08, Zhang09}, however, found that some systems exhibit decoherence freezing, where the transition probability becomes saturated after many modulation periods if the pulse delay is short. In these cases, the exponential form assumed in Ref.~\cite{Facchi05} is violated. Therefore, it is of interest to directly compare in the same system the transition probabilities (see Fig.~\ref{fig:gtaus}), instead of the transition rates, under either frequent measurements or frequent modulations.

In this paper, we revisit the study of the transitions of many spins in random magnetic fields under either frequent measurements or frequent modulations. The spins will be assumed to be initially in their spin-up state. Using exact numerical calculations, we systematically compare the performance of the three control methods in suppressing/enhancing the transition probabilities for three distributions (Gaussian, Lorentzian, and exponential) of the random local fields.

This paper is organized as follows. In Sec.~\ref{sec:dec} we formulate the spin system's free dynamics and controlled dynamics under either (i) frequent measurements, (ii) frequent modulations, or (iii) a mix of modulations and measurements. We present numerical results in Sec.~\ref{sec:num} for the three magnetic field distributions and compare in detail the performance of the three control methods listed above. Conclusions and discussions are presented in Sec.~\ref{sec:con}.

\section{Evolution of the system}
\label{sec:dec}

We consider $K$ spin-1/2 particles in random longitudinal ($z$) magnetic fields, but spatially uniform horizontal ($x$) fields~\cite{Slichter92}:
\begin{eqnarray}
H &=& \sum_{k=1}^K {\omega_k\over 2} \sigma_{kz} + g \sum_{k=1}^K \sigma_{kx},
\label{eq:h}
\end{eqnarray}
where $k$ is the spin index, and $\sigma_{kz}$ and $\sigma_{kx}$ are the Pauli matrices for the $k$th spin. Also, $\omega_k$ and $2g$ are the Zeeman splitting of the $k$th spin along the longitudinal and transverse direction, respectively. We assume that single-spin operations and detection are not accessible, but the ensemble ones are available, which is the case in nuclear spin experiments. We also assume that $g$ is much smaller than the typical value of $\omega_k$, i.e., $b \gg g$ with $b \equiv \left(\sum_k \omega_k^2 / K\right)^{1/2}$.

\subsection{Free evolution}

By initially setting all spins in the spin-up state, the collective transition probability to the spin-down state becomes
\begin{equation}
\label{eq:fid}
p_0(t) = \frac 1 K \sum_{k=1}^K \frac {g^2} {\Omega_k^2} \sin^2 \Omega_k t
\end{equation}
with $\Omega_k^2 = (\omega_k/2)^2+g^2$. If the distribution of $\omega_k$ is dense, we can replace the sum over $k$ with an integral over $\omega$
$$
p_0(t) \approx \int \! \! d\omega \; \rho(\omega) \frac {g^2} {\Omega^2} \sin^2 \Omega t ,
$$
where $\rho(\omega)$ is the distribution function and is normalized: $\int d\omega \; \rho(\omega) = 1$. In this short time region, $\Gamma^{-1} \ll t \ll |g|^{-1}$, with $\Gamma$ being the spectrum width, the integrand $\sin^2 \Omega t / {\Omega^2} $ sharply peaks at $\Omega \approx 0$. For a flat distribution function $\rho(\omega)$, the transition probability becomes
\begin{eqnarray}
\label{eq:fidint}
p_0(t) &\approx & g^2 \rho(0) \int \!\! d\omega \; \frac {\sin^2 \Omega t} {\Omega^2} \nonumber \\
    &=& 2\pi g^2 \rho(0) \; t.
\end{eqnarray}

From Eq.~(\ref{eq:fid}), the transition rate $\gamma_0$ to the spin-down state is given by
\begin{equation}
\label{eq:gamma}
\gamma_0 \equiv \frac{dp_0}{dt} = \frac{g^2 }{K} \sum_{k=1}^K  \frac{\sin2\Omega_k t}{\Omega_k}. \nonumber
\end{equation}
Note that this transition rate $\gamma_0$ becomes constant if $t$ is long enough (but still satisfies $t\ll |g|^{-1}$). Moreover, the transition probability $p_0(t)$ is still small for large enough $t$.
In the dense-distribution approximation, the above equation for $\gamma_0$ becomes
\begin{eqnarray}
\gamma_0 &\approx & 2\pi g^2\int \!\!  d\omega \; \rho(\omega) \delta(\omega) \nonumber \\
    &=& 2\pi g^2 \rho(0)
\end{eqnarray}
which is consistent with the result of Eq.~(\ref{eq:fidint}).

\subsection{Controlled evolution under frequent modulations}

The evolution of two-level systems under many phase-modulation control pulses has been widely investigated~\cite{Kofman01a, Kofman01, Kofman04, Agarwal01, Facchi05, Feranchuk02, Zhang07b, Zhang09, Zheng08, Ai10}. The control pulses are usually assumed to be hard, collective, and instantaneous~\cite{Viola99}, changing the phases of all spins by $\pi$. We denote such a pulse as a $Z$ pulse. The unitary transformation of spins due to a $Z$ pulse is described by the operator: $Z = \otimes_k (|\uparrow\rangle \langle \uparrow | - |\downarrow\rangle \langle \downarrow |)_k$. We have here neglected the constant $i=\sqrt{-1}$ which does not affect the conclusion. The time evolution operator after $N$ modulations is~\cite{Zhang09}
\begin{equation}
U(t=N\tau) = (ZU_0)^N = \left(
                          \begin{array}{cc}
                            U_{11} & U_{12} \\
                            -U_{21} & -U_{22} \\
                          \end{array}
                        \right)^N,
\end{equation}
where
\begin{eqnarray}
U_{11} &=& U_{22}^* = \cos\Omega_k \tau -i\frac{\omega_k}{2\Omega_k} \sin\Omega_k \tau
    \nonumber \\
U_{12} &=& U_{21} = i\frac{g}{\Omega_k} \sin\Omega_k \tau. \nonumber
\end{eqnarray}
with $\tau$ being the delay between pulses. After a straightforward simplification, we obtain the controlled-evolution of the transition probability from the spin-up state to the spin-down state at time $t= N\tau$
\begin{eqnarray}
p^\prime_{\rm mod}(t) &=& \frac 1 K \sum_{k=1}^K {g^2 \over \Omega_k^2} \sin^2\Omega_k \tau
    {\sin^2 N\lambda_k \over \sin^2\lambda_k} \nonumber \\
    &\approx & \int \!\!  d\omega \; \rho(\omega) {g^2 \over \Omega^2}
    \sin^2\Omega \tau {\sin^2 N\lambda \over \sin^2\lambda},
\label{eq:moze}
\end{eqnarray}
where $\lambda_k$ is determined by $\sin^2\lambda_k = 1-(\omega_k/2\Omega_k)^2 \sin^2 \Omega_k\tau$.

In the limit of short time delay $\tau$ between pulses, $\tau \rightarrow 0$, the modulated transition probability $p'_{\rm mod}$ in Eq.~(\ref{eq:moze}) becomes
\begin{equation}
\label{eq:df}
p^\prime_{\rm mod} \approx \frac{1}{2} p_0(\tau)
\end{equation}
with $p_0(\tau)\approx g^2 \tau^2$, the transition probability in the first delay $\tau$~\cite{Zhang09}. In this limiting case, $p'_{\rm mod}$ becomes independent of the total evolution time and decay freezing occurs.

\subsection{Controlled evolution under frequent measurements}

We now assume that the measurements of the spin system are projective and periodic. The effect of such a measurement on the spin system is described by the projection operator, ${\cal P}=\otimes_k (|\uparrow \rangle \langle \uparrow |)_k$. By including the free evolution of the system during the measurement delay $\tau$, we obtain the total evolution operator in a period for a single spin as
\begin{equation}
V \equiv {\cal P} U_0 = \left(
                          \begin{array}{cc}
                            U_{11} & U_{12} \\
                            0 & 0 \\
                          \end{array}
                        \right).
\end{equation}
Note that this evolution is nonunitary because of the measurement. It is straightforward to find that for $N$ periods the evolution operator is
\begin{equation}
V^N = \left(
        \begin{array}{cc}
          U_{11}^N & \;\;U_{12}U_{11}^{N-1} \\
          0 & 0 \\
        \end{array}
      \right).
\end{equation}
The survival probability of an initially spin-up state for the $k$th spin becomes
\begin{equation}
\label{eq:pmeas}
p_{k,s} = |U_{11}^{N}|^2 = \left[1-\left(\frac{g^2}{\Omega_k^2}\right) \sin^2 \Omega_k \tau\right]^{N}.
\end{equation}
By including all spins, the above Eq.~(\ref{eq:pmeas}) becomes
\begin{equation}
p_s(t=N\tau) = \frac 1 K \sum_{k=1}^K \left[1-\left(\frac{g^2}{\Omega_k^2}\right) \sin^2 \Omega_k \tau\right]^{N}.
\label{eq:pmeasallk}
\end{equation}
Under the dense-distribution approximation
\begin{equation}
\label{eq:surv}
p_s(t) \approx \int \!\! d\omega \; \rho(\omega) \left[1-\left(\frac{g^2}{\Omega^2}\right) \sin^2 \Omega \tau\right]^{N} .
\end{equation}
As a result, the total transition probability away from the initial spin-up state becomes
\begin{equation}
\label{eq:meze}
p^\prime_{\rm meas}(t) = 1-p_s(t).
\end{equation}

In the limit $\tau\rightarrow 0$, the transition probability in Eq.~(\ref{eq:meze}) approaches
\begin{equation}
\label{eq:meas}
p'_{\rm meas}(t) \approx \gamma_{\rm meas} t
\end{equation}
with the transition rate $\gamma_{\rm meas} = g^2\tau$. Compared to the modulated case Eq.~(\ref{eq:df}), where the transition rate is zero, Eq.~(\ref{eq:meas}) gives a {\it nonzero} transition rate $\gamma_{\rm meas}$, unless $\tau$ is exactly zero. In this sense, as long as the number of pulses is large, the transition probability under frequent measurements $p'_{\rm meas}$ would always exceed that under frequent modulations (see Fig.~\ref{fig:gtaus}).

\subsection{Controlled evolution under a mix of modulations and measurements}

By combining both frequent modulations and frequent measurements, we may utilize the advantages of both control methods. Here, a mixed cycle with a period of $2\tau$ involves a modulation followed by a measurement. The nonunitary evolution operator for the cycle becomes
\begin{equation}
PU_0ZU_0 = \left(
        \begin{array}{cc}
          U_{11}^2 - U_{12}U_{21} & \;\;U_{12}(U_{11}-U_{22}) \\
          0 & 0 \\
        \end{array}
      \right).
\end{equation}
The total survival probability of an initial spin-up state at time $t=N\tau$ is
\begin{equation}
p_s(t) = \frac 1 K \sum_{k=1}^K \left(1-\frac{\omega_k^2 g^2}{\Omega_k^4}
    \sin^4 \Omega_k\tau \right)^{N/2}. \nonumber
\end{equation}
It is easy to obtain the transition probability
\begin{eqnarray}
\label{eq:hybr}
p'_{\rm mix} &=& 1-\frac 1 K \sum_{k=1}^K \left(1-\frac{\omega_k^2
    g^2}{\Omega_k^4} \sin^4 \Omega_k\tau \right)^{N/2} \nonumber \\
    &\approx & 1-\int \!\! d\omega \; \rho(\omega) \left(1-\frac{\omega^2
    g^2}{\Omega^4} \sin^4 \Omega\tau \right)^{N/2} .
\end{eqnarray}
As seen from Eq.~(\ref{eq:hybr}), it is difficult to obtain any analytical result in this case without specific information on the distribution function $\rho(\omega)$.

In the limit case $\tau \rightarrow 0$, the transition probability becomes
\begin{equation}
    p'_{\rm mix} \approx \gamma_{\rm mix} t
\end{equation}
where $\gamma_{\rm mix} = (1/2)b^2 g^2 \tau^3$. When $t < t_c$, the mixed transition probability $p'_{\rm mix}$ is smaller than the modulated transition probability $p'_{\rm mod}$. The critical time $t_c$ is
$$
t_c = (b^2\tau)^{-1}.
$$
This advantage of the mix method, for short times, is shown in Fig.~\ref{fig:gtaus}. Of course, after many pulses, the transition rate of the mix method is nonzero, while that of the modulation method is zero. Thus, $p'_{\rm mix} > p'_{\rm mod}$ eventually when $t>t_c$.

As a summary, in Table~\ref{tabl:tp} we list the results for the short-$\tau$ limit for the three control methods .
\renewcommand{\arraystretch}{1.5}
\begin{table}
\caption{\label{tabl:tp}Transition probabilities at $t=N\tau$ for three control methods (modulations, measurements, and mix) in the limit $\tau\rightarrow 0$ and $N\gg 1$.}
\begin{tabular}{c| c |c}
  \hline
  \hline
  Modulations & Measurements & Mix \\
  \hline
  $p'_{\rm mod} = (1/2) g^2 \tau^2$ & $p'_{\rm meas} = g^2\tau^2 N$ & $p'_{\rm mix} = (1/2) b^2g^2 \tau^4 N$ \\
  \hline
  \hline
\end{tabular}
\end{table}

\section{Numerical results}
\label{sec:num}

Given an arbitrary $\tau$, we have to resort to numerical calculations in order to compare the transition probabilities in Eqs.~(\ref{eq:moze}), (\ref{eq:meze}), and (\ref{eq:hybr}), except in the limiting case $\tau\rightarrow 0$. Among many forms of distribution functions $\rho(\omega)$, we consider three popular choices: Gaussian, Lorentzian, and exponential~\cite{Kofman00, Agarwal01a, Zhang09}.

The Gaussian distribution function used here has the form:
\begin{equation}\label{eq:gaussian}
\rho(\omega) = C \exp\left[-\frac{(\omega-\omega_m)^2} {2 \, \Gamma^2}\right],
\end{equation}
where $C$ is the normalization constant, $\omega_m$ the peak position, and $\Gamma$ the spectral width. In the numerical calculations, the lower and upper cutoff frequencies used here are $-\omega_c$ and $\omega_c$, respectively, with $\omega_c = 100 \Gamma$ for the three distributions.

The Lorentzian distribution function used here has the standard form:
\begin{equation}\label{eq:lorentzian}
\rho(\omega) = \frac{C}{ (\omega-\omega_m)^2 +\Gamma^2}.
\end{equation}

The exponential distribution function used is
\begin{equation}\label{eq:exponential}
\rho(\omega) = C \exp\left[-\frac{|\omega-\omega_m|} {\Gamma}\right].
\end{equation}

There are many ways to compare the performance of the three control methods. We employ two methods to compare the transition probabilities of the spin system: (a) Fix the number $N$ of pulses and varying the pulse delay $\tau$ to investigate the dependence of the performance of the control method on the pulse delay; (b) Fix the total evolution time $t = N \tau$ by varying the number of pulses (accordingly the pulse delay $\tau$) to investigate the dependence on the number of pulses.


\begin{figure}
\includegraphics[width=3.25in]{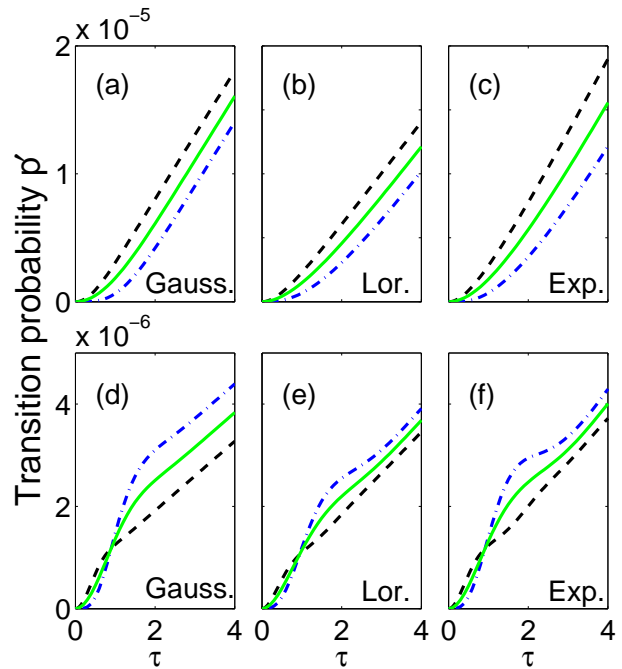}
\caption{(Color online) Transition probabilities of a many-spin system under two measurements (solid lines) or two modulations (dash-dotted lines) for three distribution functions: Left column [(a) and (d)],  Gaussian; Middle column [(b) and (e)], Lorentzian; Right column [(c) and (f)], exponential. The peaks of the distributions are chosen as $\omega_m = 0$ for the top row of panels~[(a), (b), and (c)] and $\omega_m = 2 \Gamma$ for the bottom row [(d), (e), and (f)], respectively. Dashed lines are the transition probabilities for the free evolution. The parameters $\Gamma=1$ and $g=0.001$ are used for all cases. Hereafter, time is measured in units of $\Gamma^{-1}$. The transition probabilities under two modulations are smaller than those under measurements for {\it all} $\tau$s, if $\omega_m =0$ (top panels), and for {\it small} $\tau$s, if $\omega_m = 2 \Gamma$ (bottom panels).}
\label{fig:twopulses}   
\end{figure}


\subsection{Comparison of two-pulse results with different pulse delays}

As a starting point, let us compare two-pulse effects via either modulations, measurements, or the mix of a modulation followed by a measurement. It is easy to find that the transition probability subject to two modulations is
\begin{equation}
p^\prime_{\rm mod} = \int \!\! d\omega \; \rho(\omega) \frac{g^2\omega^2} {\Omega^4} \sin^4\Omega \tau
\end{equation}
and that under two measurements
\begin{equation}
p^\prime_{\rm meas} = \int \!\! d\omega \; \rho(\omega) \frac{g^2} {\Omega^2} \sin^2\Omega \tau \left(2-\frac{g^2} {\Omega^2} \sin^2\Omega \tau\right) .
\end{equation}
The mix of one modulation and one measurement is exactly the same as two modulations, $p'_{\rm mix} = p'_{\rm mod}$.

We plot the two-pulse transition probabilities in Fig.~\ref{fig:twopulses} for Gaussian, Lorentzian, and exponential distributions. For the cases of $\omega_m = 0$, the top row of Fig.~\ref{fig:twopulses} shows that both modulations and measurements suppress the transition probability (quantum Zeno effect), compared to the free-evolution case. In addition, the transition probabilities under two modulations $p'_{\rm mod}$ are {\it always smaller} than those under two measurements $p'_{\rm meas}$, $p'_{\rm mod} < p'_{\rm meas}$. While for the cases of $\omega_m = 2 \Gamma$, both methods (i.e., two measurements or two modulations) also suppress the transition probability and $p'_{\rm mod} < p'_{\rm meas}$, if $\tau$ is small, but the two methods enhance the transition probability (quantum anti-Zeno effect) and $p'_{\rm mod} > p'_{\rm meas}$, if $\tau$ is large. The cross $p'_{\rm mod} = p'_{\rm meas}$ occurs around $\tau \approx 1$ in Fig.~\ref{fig:twopulses}. It is interesting that all three distributions show consistent and similar results.

As shown in Fig.~\ref{fig:twopulses}, for all the cases considered here, the transition probabilities increases {\it nonlinearly}, for small $\tau$s, and {\it linearly} for large $\tau$s. It is proved in Eq.~(\ref{eq:fidint}) that the transition probability increases linearly when $\omega_c^{-1} \ll t\ll g^{-1}$ for the free evolution of the spin system. Thus the nonlinearity shown for short $\tau$s indicates that the quantum Zeno and anti-Zeno effects may occur only in this short and nonlinear region. Beyond this short-$\tau$ region, the controlled evolutions are essentially the same as the free evolution.

\subsection{Comparison of multi-pulse results at fixed evolution time}

\begin{figure}
\includegraphics[width=3.25in]{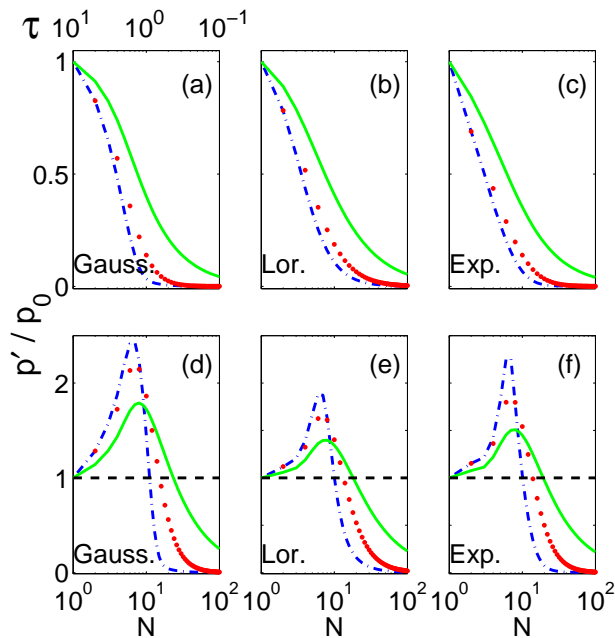}
\caption{(Color online) Quantum Zeno effect and anti-Zero effect in controlling the transition probabilities with frequent measurements (green solid lines), frequent modulations (blue dash-dotted lines), and the mix of modulations and measurements (red dotted lines). The transition probabilities subject to control pulses are normalized by dividing the free-evolution transition probability for the same time. The horizontal black dashed lines in the panels (d, e, f) of the bottom row mark the boundary between quantum Zeno and quantum anti-Zeno effect. The distribution functions are Gaussian [left column, (a) and (d)], Lorentzian [middle column, (b) and (e)], and exponential [right column, (c) and (f)]. In the top and the bottom row, $\omega_m = 0$ and $2\Gamma$, respectively. The total evolution time $t=N\tau = 10$ is fixed. Other parameters are the same as in Fig.~\ref{fig:twopulses}. In the top row, the modulation method outperforms the measurement one in suppressing the transition probability of the spin system. While in the bottom row, the modulation method is worse than the measurement one if the number of pulses $N$ is small, but better if $N$ is large.}
\label{fig:tau}   
\end{figure}

The transition probabilities of the spin system at a fixed time $t = N\tau =10$ for Gaussian, Lorentzian, and exponential distributions are shown in Fig.~\ref{fig:tau}. We normalize the transition probabilities under measurements, modulations, or the mix of modulations and measurements, by dividing the transition probability of the free evolution at the same time. In the top row of Fig.~\ref{fig:tau}, where $\omega_m = 0$, the transition probabilities under frequent modulations are smaller than those under frequent measurements and the probabilities for the mix method lie in between, i.e., $p'_{\rm mod} < p'_{\rm mix} < p'_{\rm meas}$. In addition,  $p'_{\rm mod, \;meas, \;mix} < p_0$ for all $N$, showing that frequent modulations, frequent measurements, and the mix method all suppress the transition probability and only the quantum Zeno effect occurs.

The bottom row of Fig.~\ref{fig:tau}, where $\omega_m = 2 \Gamma $, shows  a more complex phenomenon:
\begin{enumerate}
  \item For small $N$, the quantum anti-Zeno effect appears $p'_{\rm mod,\; meas,\; mix}>p_0$ (i.e., the enhancement of the transition probabilities) for all control methods.
  \item For large $N$, the quantum Zeno effect $p'_{\rm mod, \;meas, \;mix} < p_0$ (i.e., the suppression of the transition probabilities) appears for all control methods.
  \item For the same value of $N$, by comparing the performance of the modulation, the measurement, and the mix methods, we find $p'_{\rm mod} > p'_{\rm mix} > p'_{\rm meas}$ for small $N$ but $p'_{\rm mod} < p'_{\rm mix} < p'_{\rm meas}$ for large $N$.
  \item All three methods intersect around $N=10$ (or $\tau = 1$), where $p'_{\rm mod} \approx p'_{\rm meas} \approx p'_{\rm mix}$.
\end{enumerate}

For extremely small $\tau$ (large $N$), which is outside the region shown in Fig.~\ref{fig:tau}, the mix method performs better than the modulation method. For a given fixed total time $t$, we find that the critical value of $\tau$ is
$$
\tau_c = \frac{1}{b^2 t}.
$$
Correspondingly, the critical number of pulses is
$$
N_c = t/\tau_c = b^2 t^2.
$$

We remark here that the above conclusions on the general properties of the quantum Zeno and anti-Zeno effects agree with results obtained with other analytical methods in certain approximations~\cite{Kofman04, Facchi05}. The difference between ours and Ref.~\cite{Kofman04, Facchi05} is that we do not assume any specific decay form of the transition probability while an exponential form is assumed in those references. In fact, an exponential decay form is not always the case, especially with the modulation method (see Fig.~\ref{fig:gtaus}).

\subsection{Short $\tau$ limit at fixed time}
\label{sec:srt}

\begin{figure}
\includegraphics[width=3.25in]{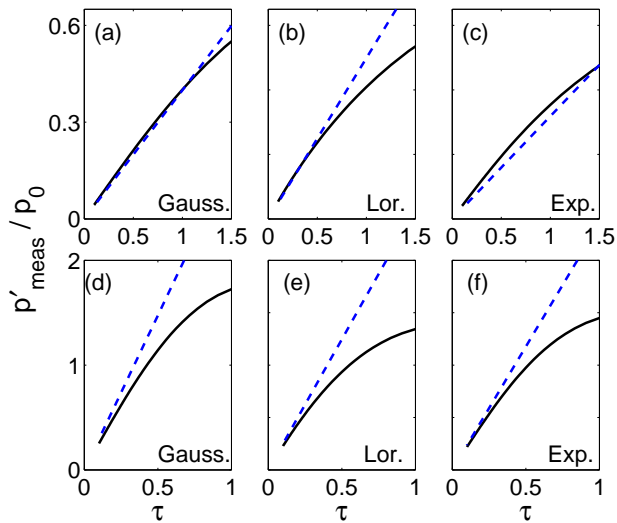}
\caption{ (Color online) Transition probability of the spin system under frequent measurements at the fixed time $t=10$ for different short pulse delays $\tau$ (black solid line). The distributions are Gaussian (left column), Lorentzian (middle column), and exponential (right column). Also, we use $\omega_m = 0$ (top row) and $\omega_m = 2\Gamma$ (bottom row). The blue dashed lines are obtained from Eq.~(\ref{eq:st}).}
\label{fig:shorttau}   
\end{figure}

Noticing that all the curves for different distribution functions in Fig.~\ref{fig:tau} give similar tails at large $N$ (small $\tau$), we next compare how the controlled system approaches its limiting case $\tau\rightarrow 0$. For the free evolution, the transition probability of the spin system can be approximated as follows if $t\ll \gamma_0^{-1}$
\begin{equation}
p_0 \approx \gamma_0 t
\end{equation}
with $\gamma_0 = 2\pi g^2 \rho(0)$ being the transition rate. Similarly, the transition probability under frequent measurements [see Eq.~(\ref{eq:meas})] at times $t \ll \gamma_{\rm meas}^{-1}$ is
\begin{equation}
p'_{\rm meas} \approx \gamma_{\rm meas} t.
\end{equation}
The above equation shows that the transition rate $\gamma_{\rm meas} = g^2\tau$ can be defined. Compared to the case of free evolution, this rate $\gamma_{\rm meas}$ is independent of the distribution function and depends linearly on the pulse delay $\tau$.

Quite differently, the transition probability [see Eq.~(\ref{eq:df})] under frequent modulations is frozen for small $\tau$,
\begin{equation}
p'_{\rm mod} \approx {1\over 2} g^2 \tau^2. \label{eq:sdf}
\end{equation}
Obviously, no decay rate can be defined. It is remarkable that this transition probability is independent of the total evolution time and the distribution function. Decoherence is frozen after several control pulses if $\tau$ is small~\cite{Zhang07a, Zhang09, Liu07}. This relationship shown in Eq.~(\ref{eq:sdf}) has been verified in Ref.~\cite{Zhang09}.

Under the mix control method, the transition probability grows linearly
\begin{equation}
p'_{\rm mix} \approx \gamma_{\rm mix} t
\end{equation}
with a well-defined transition rate $\gamma_{\rm mix} = b^2 g^2 \tau^3$ in the limit $\tau \rightarrow 0$. Note that this transition rate depends on the distribution function since $b^2 = \int \!\! d\omega \; \omega^2 \rho(\omega) $.

In the small $\tau$ limit, it is easy to obtain
\begin{eqnarray}
\label{eq:st}
\frac{p'_{\rm meas}} {p_0} &\approx & \frac{\tau}{2\pi\rho(0)}, \\
\frac{p'_{\rm mix}} {p_0} &\approx & \frac{b^2\tau^3}{2\pi\rho(0)}.\nonumber
\end{eqnarray}
We numerically check the relationship for frequent measurements by redrawing the large-$N$ results of  Fig.~\ref{fig:tau} and using $\tau$ as the horizontal axis. Figure~\ref{fig:shorttau} shows how the above limiting results are approached for different distributions as $\tau$ approaches zero.

\section{Conclusion and discussion}
\label{sec:con}

In summary, using numerical calculations, we compare the transition probabilities of a many-spin system in local random fields (with Gaussian, Lorentzian, or exponential distributions) under either frequent modulations, frequent projective measurements, or the mix of modulations and measurements. In the small-$\tau$ region, all three control methods suppress the collective transition probability of the system. Among the three control methods, the modulation one exhibits the largest suppression of the transition probability if the total evolution time is larger than the critical time, and the transition freezes after many modulation pulses. If the time is smaller than the critical time, the mix method is the most effective at suppressing the transition probabilities.

In the large $\tau$ region, all three control methods also suppress the transition probability if $\omega_m=0$, but enhance the transition probability if $\omega_m$ is large. Overall, the modulation method changes more drastically the dynamics of the system~\cite{Kofman01}: The modulation method outperforms the other two methods in either suppressing the transition in the small $\tau$ region or enhancing the transition in the large $\tau$ region, provided that the evolution time is larger than the critical time $t_c$.

The modulation-pulse delay is the same in all of our calculations. By adopting varying pulse delay, such as concatenation or Uhrig's protocol~\cite{Khodjasteh05, Uhrig07}, we could in principle suppress/enhance much more the transition probability. We believe that the transition probability would deviate more from an exponential decay under these more complicated modulation pulse sequences.

\section{Acknowledgement}

WZ is grateful for discussions with S. Ashhab, C. P. Sun, and S. Pascazio. WZ and JQY acknowledge support by the National Basic Research Program of China under Grant No. 2009CB929300. WZ acknowledges support by the National Natural Science Foundation of China under Grant No. 10904017, NCET, Specialized Research Fund for the Doctoral Program of Higher Education of China under Grant No. 20090071120013, and Shanghai Pujiang Program under Grant No. 10PJ1401300. FN was partially supported by the LPS, NSA, ARO, DARPA, AFOSR, NSF grant No. 0726909, JSPS-RFBR contract No. 09-02-92114, Grant-in-Aid for Scientific Research (S), MEXT Kakenhi on Quantum Cybernetics, and the JSPS via its FIRST program.



\end{document}